\documentclass[aps,prb,twocolumn,showpacs,amsmath,amssymb]{revtex4-1}

\usepackage[utf8]{inputenc}

\usepackage{graphicx}
\usepackage[colorlinks=true,citecolor=blue]{hyperref}
\usepackage{units}

\usepackage{verbatim} 
\usepackage{abbrevs}

\newcommand{\bra}[1]{\langle #1|}
\newcommand{\ket}[1]{|#1\rangle}
\renewcommand{\vec}[1]{\mathbf{#1}}
\newcommand{\kBT}{k_\text{B}T}

\def\up{\uparrow}
\def\down{\downarrow}

\begin{document}

\title{Electronic waiting-time distribution of a quantum-dot spin valve}

\author{Björn Sothmann}
\affiliation{Département de Physique Théorique, Université de Genève, CH-1211 Genève 4, Switzerland}

\date{\today}

\begin{abstract}
We discuss the electronic waiting-time distribution of a quantum-dot spin valve, i.e. a single-level quantum dot coupled to two ferromagnetic electrodes with magnetizations that can point in arbitrary directions. We demonstrate that the rich transport physics of this setup such as dynamical channel blockade and spin precession in an interaction-driven exchange field shows up in the waiting-time distribution and analyze the conditions necessary to observe the various effects.
\end{abstract}



\maketitle

\section{\label{sec:Intro}Introduction}
\begin{figure}
	\centering\includegraphics[width=\columnwidth]{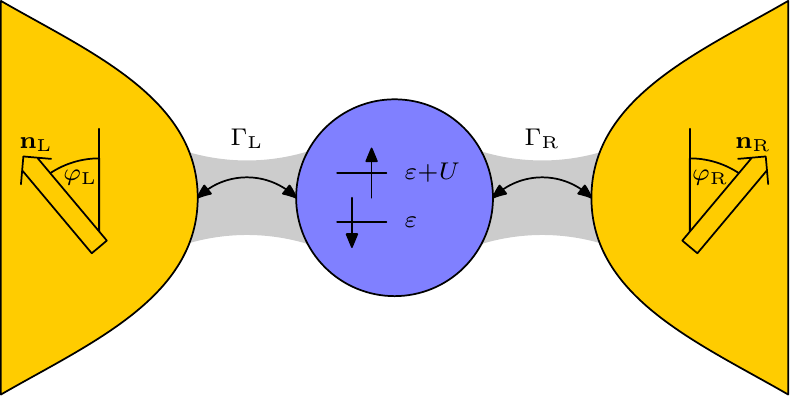}
	\caption{\label{fig:model}Schematic sketch of a quantum-dot spin valve. A single-level quantum dot (blue) with excitation energies $\varepsilon$ and $\varepsilon+U$ is tunnel coupled with coupling strength $\Gamma_r$ to two ferromagnetic electrodes (yellow) with noncollinear magnetizations pointing along $\vec n_r$ and enclosing an angle $\varphi_\text{L}+\varphi_\text{R}$.}
\end{figure}

Spin-dependent transport through nanostructures such as quantum dots has recently created a lot interest due to potential spintronics applications.~\cite{fabian_semiconductor_2007,barnas_spin_2008} Of particular interest are quantum-dot spin valves~\cite{konig_interaction-driven_2003,braun_theory_2004,braig_rate_2005} that consist of a single-level quantum dot tunnel coupled to ferromagnetic electrodes with magnetizations pointing in arbitrary directions, cf. Fig.~\ref{fig:model}. On the one hand, these systems show a  spin accumulation due to spin-dependent tunneling that has the tendency to block transport through the device. On the other hand, there is spin precession in an energy-dependent exchange field generated by virtual tunneling between the dot and the leads that lifts this blockade. The interplay between these two effects gives rise to a number of distinctive transport signatures such as a broad area of negative differential conductance,~\cite{braun_theory_2004,sothmann_probing_2010} characteristic features in the finite-frequency noise at the Larmor frequency associated with the spin precession,~\cite{braun_frequency-dependent_2006,sothmann_influence_2010,sothmann_transport_2010} a splitting of the Kondo resonance~\cite{martinek_kondo_2003,martinek_kondo_2003-1,martinek_gate-controlled_2005,utsumi_nonequilibrium_2005,sindel_kondo_2007} and a nonequilibrium spin-precession resonance.~\cite{hell_spin_2014} Other studies of such systems investigated the dependence of the current on the angle between the magnetizations,~\cite{fransson_angular_2005,pedersen_noncollinear_2005,fransson_angular_2005-1,weymann_cotunneling_2005,weymann_cotunneling_2007} the full counting statistics of electron transport,~\cite{lindebaum_spin-induced_2009} adiabatic pumping~\cite{splettstoesser_adiabatic_2008} and the possibility to generate a spin accumulation in a thermoelectric fashion.~\cite{muralidharan_thermoelectric_2013}
Experimentally, quantum dots coupled to ferromagnetic electrodes have been realized in a number of different ways, e.g., by using metallic nanoparticles,~\cite{deshmukh_using_2002,bernand-mantel_evidence_2006,wei_saturation_2007,mitani_current-induced_2008,bernand-mantel_anisotropic_2009,birk_spin-polarized_2009,birk_magnetoresistance_2010,bernand-mantel_anisotropic_2011} quantum dots defined in semiconductor nanowires,~\cite{hofstetter_ferromagnetic_2010} carbon nanotubes,~\cite{jensen_hybrid_2004,jensen_magnetoresistance_2005,sahoo_electric_2005,liu_spin-dependent_2006,hauptmann_electric-field-controlled_2008,merchant_current_2009,aurich_permalloy-based_2010,feuillet-palma_conserved_2010,gaass_universality_2011} self-assembled semiconductor quantum dots~\cite{hamaya_spin_2007,hamaya_electric-field_2007,hamaya_kondo_2007,hamaya_oscillatory_2008,hamaya_tunneling_2008,hamaya_spin-related_2009} and even single molecules.~\cite{pasupathy_kondo_2004,yoshida_gate-tunable_2013}

The investigation of electronic waiting times in transport through nanostructures is another field that recently generated a lot of interest. Waiting-time distributions have been studied for systems that can be described by generalized master equations,~\cite{koch_full_2005,brandes_waiting_2008,welack_waiting_2009,albert_distributions_2011,rajabi_waiting_2013,thomas_electron_2013} scattering matrix theory~\cite{albert_electron_2012,dasenbrook_floquet_2014,albert_waiting_2014,haack_distributions_2014} as well as in terms of noninteracting tight-binding chains.~\cite{thomas_waiting_2014} Waiting times were shown to provide information about the short-time behaviour of transport processes that cannot be obtained from other quantities such as the zero-frequency current noise or the full counting statistics.~\cite{albert_electron_2012} Furthermore, waiting times contain information about the coherent internal dynamics of quantum systems~\cite{brandes_waiting_2008,welack_waiting_2009,rajabi_waiting_2013,thomas_electron_2013} and can serve to characterize recently developed single-electron sources.~\cite{albert_distributions_2011,dasenbrook_floquet_2014,albert_waiting_2014}

Here, we study the waiting-time distribution of electron transport through a quantum-dot spin valve. We focus on the regime where sequential tunneling dominates transport and the system can be described in terms of a generalized master equation with transition rates obtained via a real-time diagrammatic approach.~\cite{konig_zero-bias_1996,konig_resonant_1996,braun_theory_2004} Our main aim is to demonstrate how the rich transport physics of the quantum-dot spin valve can be detected in the waiting-time distribution. In addition, we want to elucidate the conditions that need to be fulfilled in order to observe a specific transport feature in the waiting-time distribution and compare to the corresponding conditions required to observe the same feature in other transport properties such as zero- and finite-frequency noise or the full counting statistics.

The paper is organized as follows. In Sec.~\ref{sec:Model} we introduce the model of a quantum-dot spin valve. Section~\ref{sec:Theory} describes the theoretical approach to calculate the waiting-time distribution within the framework of a real-time diagrammatic approach. We present our results in Sec.~\ref{sec:Results} and conclude with a summary in Sec.~\ref{sec:Summary}.

\section{\label{sec:Model}Model}

We consider a single-level quantum dot coupled to two ferromagnetic electrodes $r=\text{L,R}$ with magnetizations pointing in arbitrary directions $\vec n_r$. The Hamiltonian of the system can be written as
\begin{equation}
	H=\sum_r H_r+H_\text{dot}+H_\text{tun}.
\end{equation}
Here,
\begin{equation}
	H_r=\sum_{\vec k\sigma}\varepsilon_{r\vec k\sigma}a_{r\vec k\sigma}^\dagger a_{r\vec k\sigma}
\end{equation}
represents the two ferromagnetic electrodes in terms of a simple Stoner model as noninteracting electrons with a constant but spin-dependent density of states $\rho_{r\sigma}$. For each electrode, we choose the spin quantization axis along the magnetization of the respective lead such that $\sigma=\pm$ refers to the majority (minority) spin electrons. The spin dependence of $\rho_{r\sigma}$ can be conveniently parametrized in terms of the polarization $p_r=(\rho_{r+}-\rho_{r-})/(\rho_{r+}+\rho_{r-})$ where $p_r=0$ refers to a normal metal and $p_r=1$ to a half-metallic ferromagnet. In the following, we will assume both leads to have the same polarization, $p_\text{L}=p_\text{R}\equiv p$.

The quantum dot is described in terms of a single spin-degenerate level with gate-tunable energy $\varepsilon$ as
\begin{equation}
	H_\text{dot}=\sum_\sigma \varepsilon c_\sigma^\dagger c_\sigma +Uc_\up^\dagger c_\up c_\down^\dagger c_\down.
\end{equation}
Here, $U$ denotes the Coulomb energy of the dot that is needed in order to occupy the quantum dot with two electrons at the same time. For later convenience, we quantize the dot spin along the direction $\vec n_\text{L}\times\vec n_\text{R}$ perpendicular to the lead magnetizations.

For the chosen quantization axes, the tunnel Hamiltonian takes the form 
\begin{align}\label{eq:Htun}
\begin{split}
	H_\text{tun}=\sum_{r\vec k}\frac{t_r}{\sqrt{2}}\left[a_{r\vec k+}^\dagger(e^{i\varphi_r/2} c_\up+e^{-i\varphi_r/2}c_\down)\right.\\
	\left.+a_{r\vec k-}^\dagger(-e^{i\varphi_r/2} c_\up+e^{i\varphi_r/2}c_\down)\right]+\text{H.c.},
\end{split}
\end{align}
i.e., it couples majority and minority spin electrons of the lead to both, spin up and spin down electrons on the dot. In Eq.~\eqref{eq:Htun}, $\varphi_r$ denotes the angle between $\vec n_r$ and $\vec n_\text{L}+\vec n_\text{R}$. The tunnel matrix elements $t_r$ are related to the spin-dependent tunnel coupling strengths via $\Gamma_{r\sigma}=2\pi|t_r|^2\rho_{r\sigma}$ with $\Gamma_r=(\Gamma_{r+}+\Gamma_{r-})/2$. We furthermore introduce the total tunnel coupling $\Gamma=\Gamma_\text{L}+\Gamma_\text{R}$.

\section{\label{sec:Theory}Theory}
In order to describe transport through our system, we employ a real-time diagrammatic technique~\cite{konig_zero-bias_1996,konig_resonant_1996} in its extension to systems with ferromagnetic leads.~\cite{braun_theory_2004,braun_frequency-dependent_2006} The central idea of this approach is to split the system into the strongly interacting quantum dot with a few degrees of freedom and the noninteracting electrodes with many degrees of freedom. The latter are integrated out to obtain a description of the quantum dot in terms of its reduced density matrix $\rho^\text{red}$ with density matrix elements $P^{\chi_1}_{\chi_2}=\bra{\chi}\rho^\text{red}\ket{\chi}$. The time evolution of the reduced density matrix is given by a generalized master equation of the form $\dot {\vec P}=\vec W\vec P$. Here, $\vec P$ is a vector containing the various density matrix elements of $\rho^\text{red}$. $\vec W$ is a matrix of generalized transition rates that are given by irreducible self-energies of the quantum-dot propagator on the Keldysh contour. They can be evaluated in a systematic expansion in the tunnel couplings while taking into account interaction and nonequilibrium effects exactly.~\cite{braun_theory_2004,braun_frequency-dependent_2006} In the following, we restrict ourselves to first order terms only which is a good approximation as long as $\Gamma_r\ll\kBT$.

For the quantum-dot spin valve, we can rewrite the generalized master equation in a physically intuitive form by introducing the probabilities to find the dot empty $P_0$, singly occupied $P_1=P_\up+P_\down$ or doubly occupied $P_d$ as well as the quantum-statistical average of the dot spin $S_x=(P^\up_\down-P^\down_\up)/2$, $S_y=(P^\up_\down-P^\down_\up)/(2i)$ and $S_z=(P_\up-P_\down)/2$. The generalized master equation can then be split into one set of equations for the occupation probabilities and another one for the average spin.~\cite{braun_theory_2004} The first set of equations is given by
\begin{widetext}
\begin{equation}
	\left(
	\begin{array}{c}
		\dot P_0 \\
		\dot P_1 \\
		\dot P_d
	\end{array}
	\right)
	=
	\sum_r \Gamma_r
	\left(
	\begin{array}{ccc}
		-2f_r^+(\varepsilon) & f_r^-(\varepsilon) & 0 \\
		2f_r^+(\varepsilon) & -f_r^-(\varepsilon)-f_r^+(\varepsilon+U) & 2f_r^-(\varepsilon+U) \\
		0 & f_r^+(\varepsilon+U) & -2f_r^-(\varepsilon+U)
	\end{array}
	\right)
	\left(
	\begin{array}{c}
		P_0 \\
		P_1 \\
		P_d
	\end{array}
	\right)
	+\sum_r 2p\Gamma_r
	\left(
	\begin{array}{c}
		f_r^-(\varepsilon) \\
		-f_r^-(\varepsilon)+f_r^+(\varepsilon+U) \\
		-f_r^+(\varepsilon+U)
	\end{array}
	\right)
	\vec S\cdot \vec n_r
\end{equation}
\end{widetext}
with the Fermi function $f_r^+(\omega)=1-f_r^-(\omega)=1/\{\exp[(\omega-V_r)/\kBT]+1\}$, where $V_r$ denotes the voltage applied to lead $r$ and $T$ is the electrode temperature, assumed to be equal for both leads. Due to the ferromagnetic electrodes, the occupation probabilities not only couple to each other but also couple to the spin accumulation on the dot. The equation governing  the spin dynamics reads
\begin{equation}
	\left(\frac{d\vec S}{dt}\right)=\left(\frac{d\vec S}{dt}\right)_\text{acc}+\left(\frac{d\vec S}{dt}\right)_\text{rel}+\left(\frac{d\vec S}{dt}\right)_\text{prec}
\end{equation}
where
\begin{multline}
	\left(\frac{d\vec S}{dt}\right)_\text{acc}=\sum_rp\vec n_r\Gamma_r\left[f_r^+(\varepsilon)P_0+\frac{f_r^+(\varepsilon+U)-f_r^-(\varepsilon)}{2}P_1\right.\\\left.-f_r^-(\varepsilon+U)P_d\vphantom{\frac{1}{2}}\right]
\end{multline}
describes the accumulation of spin on the dot due to spin-polarized tunneling on and off the dot. Similarly,
\begin{equation}
	\left(\frac{d\vec S}{dt}\right)_\text{rel}=-\sum_r\Gamma_r\left[f_r^-(\varepsilon)+f_r^+(\varepsilon+U)\right]\vec S
\end{equation}
represents the relaxation of the dot spin due to electron tunneling. Finally,
\begin{equation}
	\left(\frac{d\vec S}{dt}\right)_\text{prec}=\sum_r \vec B_r\times \vec S
\end{equation}
characterizes a precession of the dot spin in the effective exchange field
\begin{equation}
	\vec B_r=\frac{p\Gamma_r\vec n_r}{\pi}\int' d\omega \left(\frac{f_r^-(\omega)}{\omega-\varepsilon-U}+\frac{f_r^+(\omega)}{\omega-\varepsilon}\right)
\end{equation}
generated by spin-dependent virtual tunneling between the dot and the electrodes.

We now detail how to calculate the waiting-time distribution between tunneling events. In particular, we focus on tunneling out of the dot into the right lead. The central object in the calculation of the waiting-time distribution is the matrix $\vec W^X$ that can be obtained from $\vec W$ in a straightforward way: One simply multiplies each diagram encountered in the evaluation of $\vec W$ with a factor +1 if the corresponding process transfers an electron from the dot into the right lead and with a factor of 0 otherwise. The waiting-time distribution is then given by~\cite{brandes_waiting_2008}
\begin{equation}
	w(\tau)=\frac{\vec e^T \vec W^X\exp[(\vec W-\vec W^X)\tau]\vec W^X \rho^\text{stat}}{\vec e^T\vec W^X \rho^\text{stat}},
\end{equation}
where $\rho^\text{stat}$ is the stationary density matrix satisfying $0=\vec W\rho^\text{stat}$ and $\vec e^T$ is a vector that picks out the diagonal density matrix elements.

In general, it is not possible to obtain compact analytical expressions for the waiting-time distribution of a quantum-dot spin valve. However, in the limiting case that transport occurs through the empty and singly-occupied dot only, $f_\text{L}^+(\varepsilon)=f_\text{R}^-(\varepsilon)=1$ and $f_\text{L}^+(\varepsilon+U)=f_\text{R}^+(\varepsilon+U)=0$, and the quantum dot is symmetrically coupled to the electrodes, $\Gamma_\text{L}=\Gamma_\text{R}=\Gamma/2$, we find for parallel magnetizations
\begin{multline}
	w_P(\tau)=\frac{(1-p)^2}{1+p}\Gamma e^{-(1-p)\Gamma\tau}+\frac{(1+p)^2}{1-p}\Gamma e^{-(1+p)\Gamma\tau}\\
	-2\frac{1+3p^2}{1-p^2}\Gamma e^{-2\Gamma\tau},
\end{multline}
antiparallel magnetizations
\begin{multline}
	w_{AP}(\tau)=(1-p)\Gamma e^{-(1-p)\Gamma\tau}+(1+p)\Gamma e^{-(1+p)\Gamma\tau}\\
	-2\Gamma e^{-2\Gamma\tau},
\end{multline}
and arbitrarily oriented magnetizations, neglecting the exchange field in the calculation,
\begin{multline}
	w_\varphi^{B_r=0}(\tau)=\frac{(1-p)(1-p\cos\varphi)}{1+p}\Gamma e^{-(1-p)\Gamma\tau}\\
	+\frac{(1+p)(1+p\cos\varphi)}{1-p}\Gamma e^{-(1+p)\Gamma\tau}\\
	-2\frac{1+p^2(1+2\cos\varphi)}{1-p^2}\Gamma e^{-2\Gamma\tau}.
\end{multline}
In all three cases, the first term arises from the tunneling out of majority spin electrons into the drain. Similarly, the second term is due to the tunneling out of minority spin electrons. The last term describes tunneling in of electrons from the source. As both majority and minority spin electrons can tunnel into the empty dot, the time scale for this exponential decay does not depend on the polarization.

\section{\label{sec:Results}Results}
In the following, we analyze the waiting-time distribution for three different transport regimes that illustrate characteristic transport features of the quantum-dot spin valve. First, we will consider electron bunching due to a dynamical channel blockade~\cite{cottet_positive_2004-1,cottet_positive_2004,cottet_dynamical_2004,belzig_full_2005,elste_transport_2006,urban_tunable_2009} which occurs for parallely magnetized electrodes.~\cite{bulka_current_2000,braun_frequency-dependent_2006} Second, we show how the spin precession in the exchange field that occurs for noncollinear magnetizations~\cite{konig_interaction-driven_2003,braun_theory_2004,braun_frequency-dependent_2006} manifests itself in the waiting-time distribution. Finally, we discuss signatures of a recently discovered dynamical spin resonance that occurs for nearly antiparallel magnetizations.~\cite{hell_spin_2014}

\subsection{Electron bunching}

\begin{figure}
	\includegraphics[width=\columnwidth]{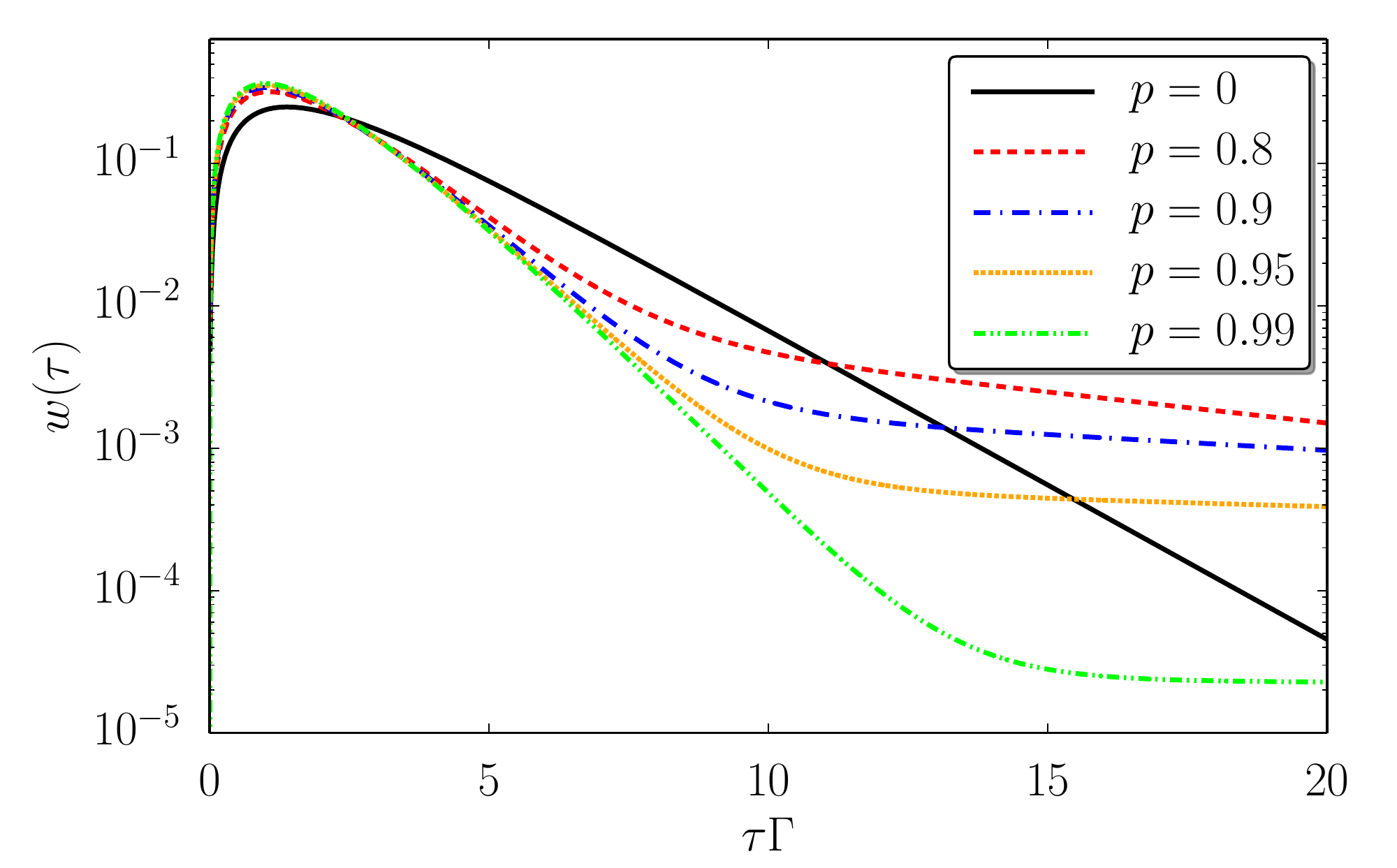}
	\caption{\label{fig:Parallel}Waiting-time distribution for a symmetric quantum-dot spin valve, $\Gamma_\text{L}= \Gamma_\text{R}=\Gamma/2$, in the parallel geometry, $\varphi_r=0$. The bias voltage is chosen such that transport occurs through the empty and singly-occupied state only. For large polarizations a crossover between two different exponential decay indicates bunching of electron transport.}
\end{figure}

For parallel magnetizations, the average charge current is independent of the polarization $p$ because as the current contribution of majority spin electrons increases with $p$, the current contribution of minority spin electrons decreases with $p$ by the same amount. However, as $p$ increases, the current becomes less regular due to electron bunching. While majority spin electrons can tunnel easily on and off the dot, the corresponding rates for minority spin electrons are suppressed by $1-p$. Thus, minority spin electrons dynamically block transport and chop the current into bunches of majority spin electrons flowing through the dot.

In the waiting time distribution, this dynamical channel blockade shows up as the crossover between two exponential decays with two different time scales. At short times, the exponential decay of the waiting-time distribution is dominated by the tunneling of majority spin electrons $e^{-(1+p)\Gamma\tau}$~\footnote{The contribution from tunneling in of electrons decays even faster as $e^{2\Gamma\tau}$. However, its coefficient is always smaller than that of $e^{-(1+p)\Gamma\tau}$ such that it never dominates the exponential decay of the waiting-time distribution.}. At longer times, the contribution due to the tunneling of minority spin electrons, $e^{-(1-p)\Gamma\tau}$, that describes the occasional blockade of the dot, takes over.

As can be seen in Fig.~\ref{fig:Parallel}, the crossover time increases as the polarization $p$ is increased. Furthermore, the crossover becomes more pronounced as $p$ grows because of the larger difference between the decay rates $(1+p)\Gamma$ and $(1-p)\Gamma$. In addition, as $p$ is increased the crossover occurs at smaller values of $w(\tau)$ which renders an experimental observation more challenging as  it requires a high statistics of tunneling events.
These findings are in agreement with the picture of bunching due to dynamical channel blockade. As $p$ grows, the blockade events due to minority spin electrons on the dot become more and more rare. In addition, the average duration of an individual blockade event becomes longer as well due to the reduced tunneling out rate of minority spins.

We finally compare our results to the signatures of the dynamical channel blockade in the zero-frequency noise~\cite{braun_frequency-dependent_2006}. In the regime where transport occurs through the empty and singly occupied dot only, the zero-frequency Fano factor, i.e. the ratio between the current noise and the average current,
\begin{equation}
	F=\frac{4(1+p^2)\Gamma_\text{L}+(1-p^2)\Gamma_\text{R}}{(1-p^2)(2\Gamma_\text{L}+\Gamma_\text{R})^2}
\end{equation}
becomes super-Poissonian, $F>1$, for polarizations larger than~\cite{sothmann_mesoscopic_2012}
\begin{equation}
	p^*=\sqrt{\frac{\Gamma_\text{R}}{2\Gamma_\text{L}+\Gamma_\text{R}}}.
\end{equation}
The Fano factor characterizes the average number of electrons within one bunch transferred through the dot. We remark that in general a larger polarization is needed to observe the electron bunching in the waiting-time distribution than in the Fano factor.

\subsection{Spin precession}
We now turn to the case of a quantum-dot spin valve with noncollinearly magnetized leads. In this situation, the dot spin precesses in an exchange field due to virtual tunneling between the dot and the leads. This spin precession gives rise to characteristic features in the finite-frequency noise at the Larmor frequency of the exchange field~\cite{braun_frequency-dependent_2006,sothmann_transport_2010}. In the following, we demonstrate how the spin precession manifests itself in the waiting-time distribution.

\begin{figure}
	\includegraphics[width=\columnwidth]{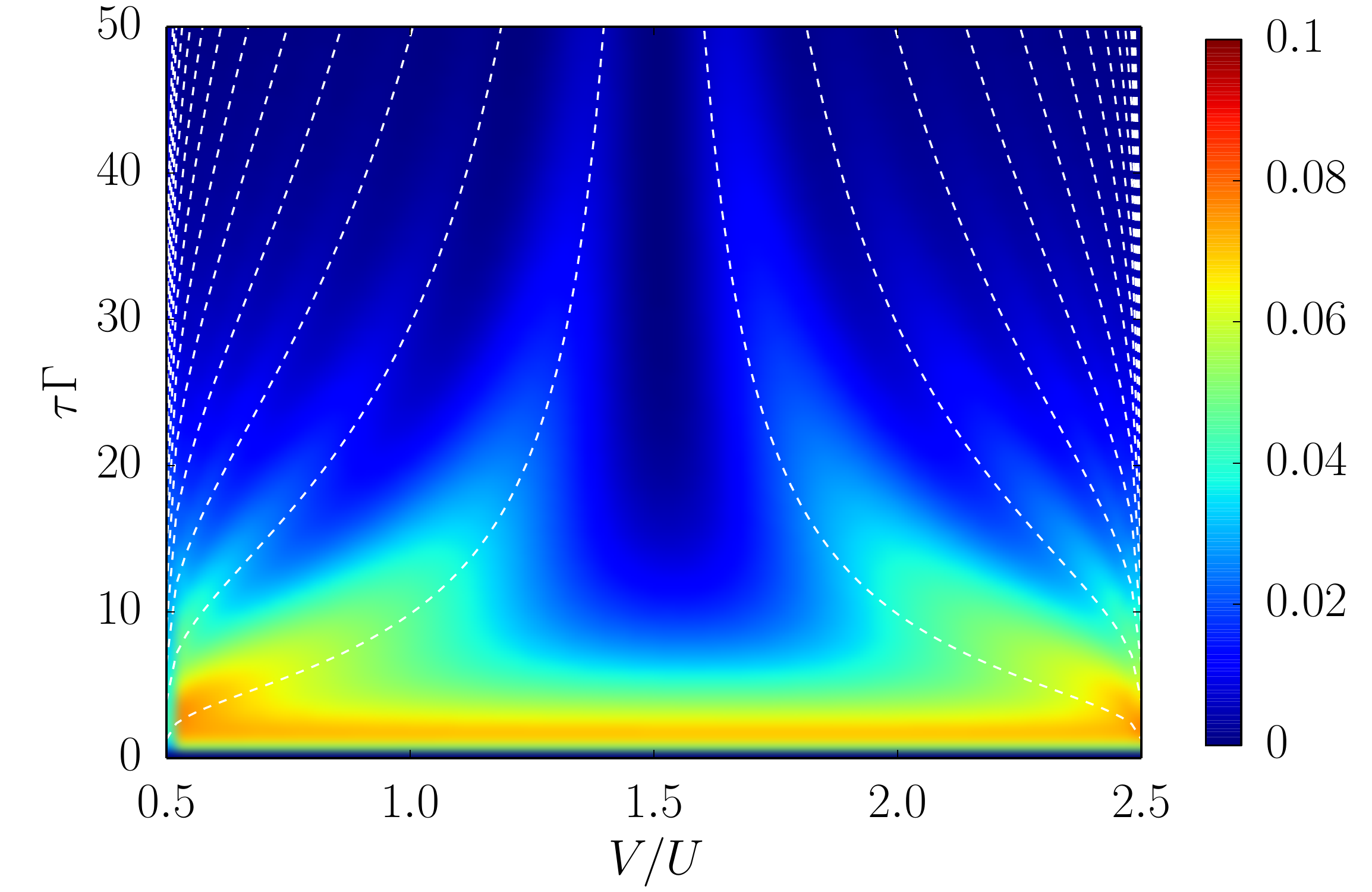}
	\caption{\label{fig:Perpendicular1}Waiting-time distribution as a function of bias voltage for a quantum-dot spin valve with perpendicular magnetizations, $\varphi_\text{L}=-\varphi_\text{R}=\pi/4$. Parameters are $\Gamma_\text{L}=10\Gamma_\text{R}$, $\varepsilon=U/4$, $p=1$. White dashed lines mark the times after which the spin of the dot has precesssed by an angle $(2n+1)\pi$.}
\end{figure}

Figure~\ref{fig:Perpendicular1} shows the waiting-time distribution as a function of the bias voltage for perpendicular magnetizations. For a given voltage, the waiting-time distribution shows oscillations with a frequency given by the Larmor frequency of the spin precession. 
The mechanism that gives rise to these oscillations is the following.
Electrons preferrably enter the dot with a spin pointing along the magnetization of the source electrode. On the dot, the electron spin precesses in the exchange field. After a precession by an angle of $(2n+1)\pi$, the overlap between the dot spin and the majority spins of the drain electrode becomes maximal. Hence, there is an increased probability to tunnel out of the dot that leads to local maxima of the waiting-time distribution. We remark that the position of the maxima is not precisely given by $\omega\tau=(2n+1)\pi$ as it takes a finite time for an electron to tunnel into the empty dot.
As the Larmor frequency depends on the energy-dependent exchange field, the frequency of the oscillations changes with bias voltage as can be clearly seen in Fig.~\ref{fig:Perpendicular1}. At the particle-hole symmetric point, $V=2\varepsilon+U$, the exchange field does not affect the dot spin at all and hence the oscillations vanish at this point.

\begin{figure}
	\includegraphics[width=\columnwidth]{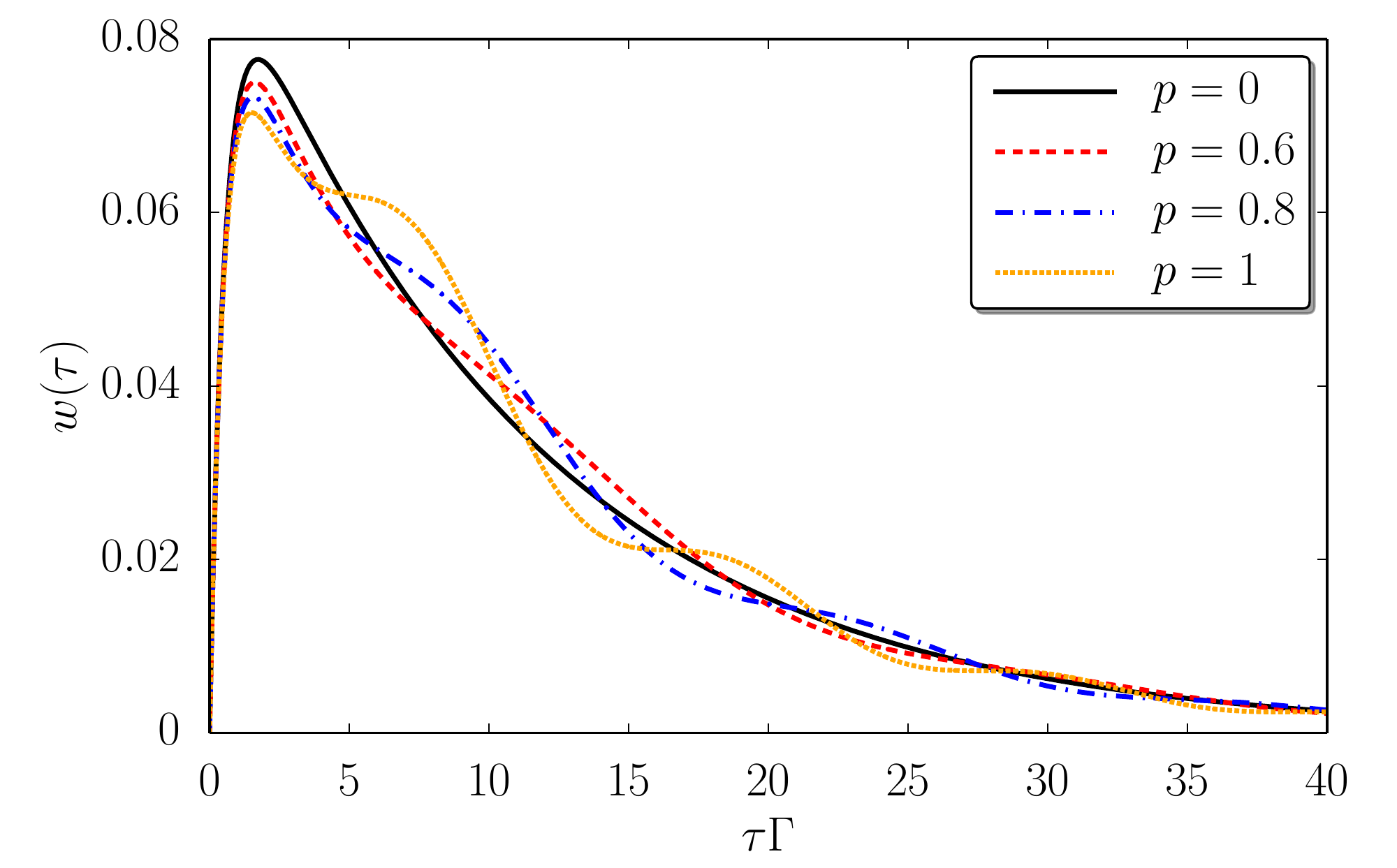}
	\caption{\label{fig:Perpendicular2}Waiting-time distribution for a quantum dot spin-valve with perpendicular magnetizations. $V=3U/4$, other parameters as in Fig.~\ref{fig:Perpendicular1}.}
\end{figure}

In Fig.~\ref{fig:Perpendicular2}, we plot the waiting-time distribution for a fixed bias voltage and different values of the polarization. While for halfmetallic leads the oscillations in the waiting times are well pronounced, they quickly get washed out as the polarization is reduced and disappear around $p=0.6$ because of the strong decoherence due to tunneling events. Hence, in order to observe the spin precession in the waiting times, highly polarized electrodes are needed. This in in contrast to the signatures in the finite-frequency noise that are observable even for moderate polarizations of $p\approx 0.3$ achieveable, e.g., with electrodes made from Fe, Ni or Co~\cite{monsma_spin_2000}.

\subsection{Spin resonance}
\begin{figure}
	\includegraphics[width=\columnwidth]{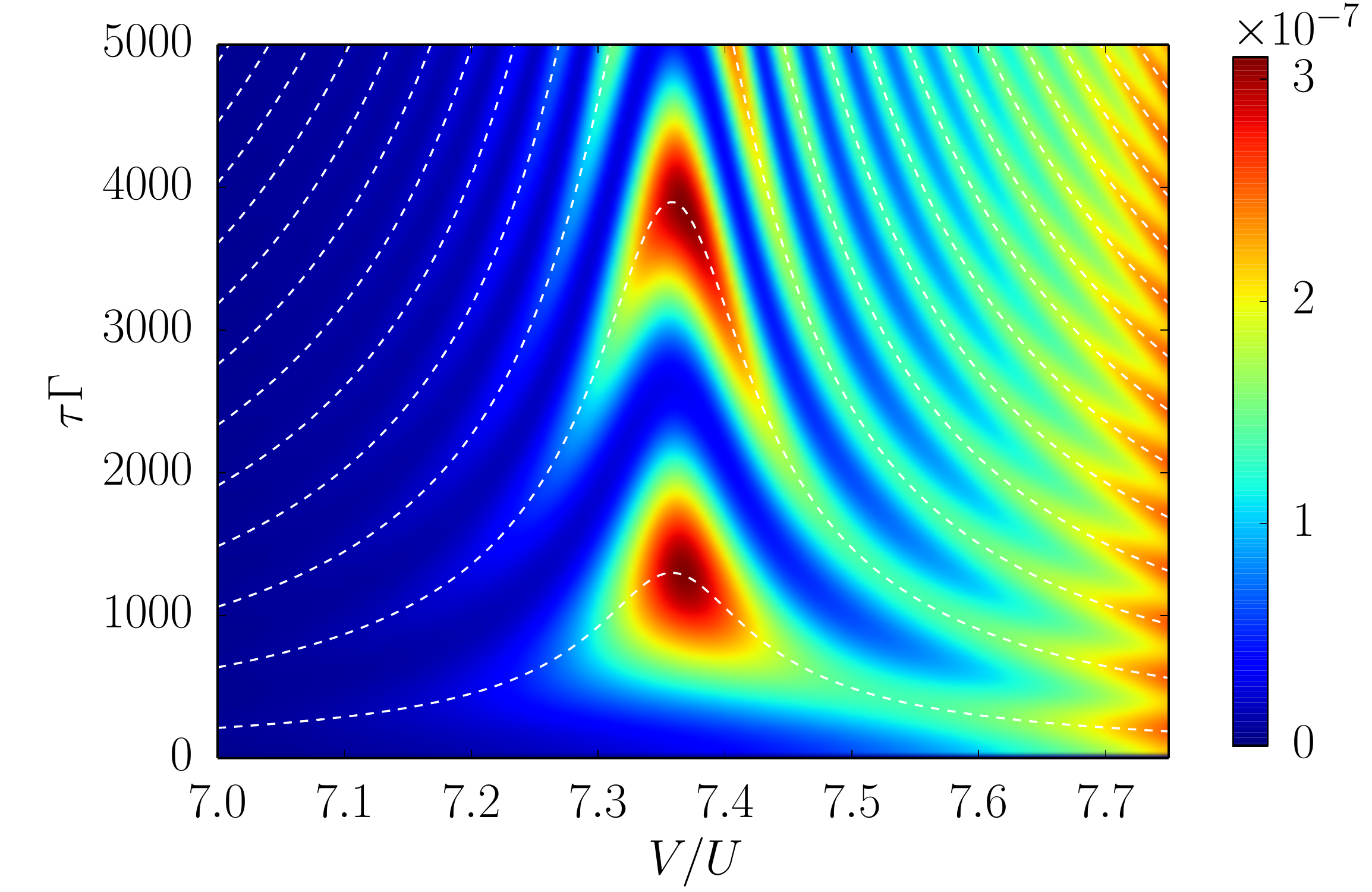}
	\caption{\label{fig:SpinResonance}Waiting-time distribution as a function of applied bias voltage and time. Parameters are $\Gamma_L=10\Gamma_R$, $\varepsilon=-U/4$, $p=1$ and $\varphi_L=-\varphi_R=0.495\pi$. White dashed lines mark again the times after which the spin has precessed by an angle $(2n+1)\pi$ in the exchange field.}
\end{figure}

As a last transport feature, we discuss how a recently predicted spin resonance~\cite{hell_spin_2014} manifests itself in the waiting-time distribution. To this end, we consider transport inside the Coulomb-blockade region where the dot is preferrably singly occupied. In principle, one needs to perform a second-order calculation here. However, we expect that our sequential tunneling approximation still captures the essential features of the spin resonance and its signatures in the waiting-time distribution. As in the Coulomb-blockade regime adding or removing an extra electron from the dot is energetically unfavorable, electrons on the dot have a long life-time such that the waiting time decays exponentially on a very long time scale.

We now focus on the case of nearly antiparallely magnetized leads. In this situation, we get on top of the Coulomb-blockade also a spin blockade as majority spin electrons from the source are minority spin electrons in the drain and hence have a suppressed probability of leaving the dot. In the nearly antiparallel geometry, the exchange field typically has a large component along the spin accumulation on the dot and only a small component perpendicular to it. Hence, there is only a very weak precession of the dot spin that cannot lift the spin blockade. In consequence, the waiting-time distribution shows a slow exponential decay with weak oscillations superimposed, cf. Fig~\ref{fig:SpinResonance} at $V=7.75 U$.

However, as the exchange field contributions from the left and right lead have indepenent dependencies on the level position and bias voltage, it is possible to achieve that the exchange field component along the spin accumulation vanishes. In this case, the spin precesses in the small remaining exchange field perpendicular to the spin accumulation. This precession periodically lifts the spin blockade of the dot and increases the current from its suppressed value in the antiparallel geometry back to value in the parallel geometry~\cite{hell_spin_2014}. In the waiting time distribution, the resonance shows up as strong oscillations with the Larmor frequency on top of a slow exponential decay, cf. Fig~\ref{fig:SpinResonance} at $V=7.35 U$. Hence, as the system is tuned through the resonance, the current exhibits a peak (not shown) while the waiting-time distribution shows an increase of the period and amplitude of its oscillations.

\section{\label{sec:Summary}Summary}
We have analyzed the electronic waiting time of a quantum-dot spin valve. We obtained analytical results for the waiting-time distribution for collinear magnetizations as well as in the noncollinear case when neglecting the effect of a tunneling induced exchange field. We then discussed signatures of characteristic transport features of a quantum-dot spin valve in the waiting-time distribution. In particular, we showed that the electron bunching due to dynamical channel blockade leads to two different exponential decays of the waiting-time distribution. The spin precession for noncollinear setups gives rise to characteristic oscillations of the waiting-time distribution. Finally, we demonstrated that a recently predicted spin resonance in the Coulomb-blockade regime gives rise to pronounced oscillations in the waiting-time distribution on a very long time scale.

\acknowledgments
I thank Christian Flindt for valuable feedback on the manuscript and acknowledge financial support from the Swiss NSF via the NCCR QSIT.


%

\end{document}